\documentclass[showpacs,preprintnumbers,amsmath,amssymb,nofootinbib,groupedaddress,superscriptaddress,showkeys]{revtex4}

\usepackage{graphicx}
\usepackage{dcolumn}
\usepackage{epsfig}
\usepackage[dvips]{color}
\usepackage{bm}

\allowdisplaybreaks[4]

\begin{document}

\preprint{KYUSHU-HET 61}
\preprint{STUPP-02-170}

\title{%
Yet another correlation in the analysis of CP violation
using a neutrino oscillation experiment}

\author{Toshihiko Ota}
\email{toshi@higgs.phys.kyushu-u.ac.jp}
\affiliation{Department of Physics, Kyushu University,\\ 
         Hakozaki, Higashi-ku, Fukuoka 812-8581, Japan}

\author{Joe Sato}
\email{joe@phy.saitama-u.ac.jp}
\affiliation{Department of Physics, Faculty of Science,\\
Saitama University, Saitama, 338-8570, Japan}


\begin{abstract}
 We investigate the effect induced by variations in the density
profile of the Earth's interior using a long-baseline neutrino oscillation
experiment.

 At first, we point out two facts.
(i) The most essential part of the matter profile is the
first Fourier mode of the matter profile function;
and (ii) The Earth models based on the study of seismology include a large
uncertainty for the first Fourier mode.
 Next, we show that there is a strong correlation between the average
density value and the first Fourier coefficient in the analysis of 
oscillation probability.
 This means that the matter profile effect induces added uncertainty for
the average matter parameter. 

 Taking into account this extra uncertainty, we make numerical
calculations for the sensitivity to CP violation search and 
 show that CP sensitivity is impaired by this added uncertainty
within a large $U_{e3}$ region.
\end{abstract}



\maketitle

\section{introduction}
The effects of the Earth's matter \cite{MSW} play a very important role
in long-baseline and high-energy oscillation experiments, such as a
neutrino factory \cite{nuFact}.  Furthermore, the effect of variations
in matter density and chemical composition along the baseline on the
oscillation probability, which we call the matter profile effect, is
also a controversial issue and has been debated in various contexts
\cite{Smirnov, Petcov, Akhmedov, Ohlsson, Mocioiu, Fogli, Lindner,
Takasugi, Shan, Jacobsson, reconstruction, KS, OS}.

In the search for CP-violating effects, it is important to understand the
structure of the Earth so as to estimate precisely the fake signal
induced by the Earth's matter.  Some research \cite{reconstruction} has
concluded that it is hard to obtain information about the interior of
the Earth through neutrino experiments using the currently assumed size
of statistics and realistic detector configuration.  Therefore, the
science of geophysics has been applied to support the analysis.  
Much analysis has used the Preliminary Reference Earth Model (PREM)
\cite{PREM}, an Earth model which is based on seismology.  When we deal
with such a model, we need to be conscious of the fact that the model
includes uncertainty.  In this paper, we discuss the fact that this uncertainty
reduces the sensitivity of the CP violation and we show that it is not
a small effect.

The procedure used to demonstrate our argument is as follows: Firstly, we
recapitulate the method of the Fourier series expansion of the matter
profile function \cite{KS,OS} in Sec.\ref{Sec:FourierMethod}.  Using
this method, we show that only the first few modes are important in a 
high-energy experiment.  Furthermore, we mention the uncertainty
included in the
seismological Earth models.  We point out that a few percent error for
the profile function can actually be interpreted as huge uncertainty for the
Fourier coefficients, which can affect the CP sensitivity.  Next, in
Sec.\ref{Sec:Colleration}, we show that there is a strong correlation
between the constant matter parameter and the first Fourier coefficient of
the matter profile function within a wide energy and baseline region.  This
fact means that the uncertainty of the matter profile effect brings
added uncertainty to the average matter density.  Then, we present
numerical calculations to show quantitatively a loss of sensitivity
induced by the uncertainty of the matter profile effect in 
Sec.\ref{Sec:Numerical}.
Finally, a summary is given in Sec.\ref{Sec:Summary}.

\section{Method of the Fourier series}\label{Sec:FourierMethod}
We would like to point out two facts in this section.
{\it 
(1) The most essential part of the matter profile effect is the first
    Fourier mode of the matter profile function.
(2) The seismological Earth models include great uncertainty for the first
Fourier mode.}
These facts force us to reconsider analyses in the baseline region
which have so far assumed that the matter profile effect is not
significant.  In the case where the baseline length is 3,000 km,
PREM tells us that the matter profile function is almost flat and hence
it can be expected that the matter profile effect is small.  However, if we consider
the uncertainty of PREM, how will the analysis change?

\subsection{Introduction of the method}
To see the effects induced by the matter profile, 
we derive the analytic expression using the method of the Fourier series
\cite{KS, OS}.
Expanding the matter profile function into the Fourier modes, we obtain
an extremely clear viewpoint for the resonance conditions 
between the oscillation
lengths of the neutrino and the matter profile undulation. By this
expansion we can understand which modes, and what structures, are effective%
\footnote{%
The Fourier expanded-matter profile can not reproduce the boundary
between two layers precisely.  However, it will be shown below that the
higher modes which construct the edge are not effective in high-energy
experiments.
}.

Now, we introduce our calculation method.
We assume three generations and parameterize the mixing matrix of the
lepton sector, Maki-Nakagawa-Sakata-Pontecorvo matrix (MNSP matrix), 
as follows;
\begin{align}
 U_{\alpha i} 
&=\begin{pmatrix}
     1 &    & \\
       & c_{\psi}  & s_{\psi} \\
       & -s_{\psi} & c_{\psi}
   \end{pmatrix}
   \begin{pmatrix}
     1 &   &  \\
       & 1 &  \\
       &   & e^{i \delta}
   \end{pmatrix}
   \begin{pmatrix}
     c_{\phi} &   & s_{\phi} \\
              & 1 &          \\
     -s_{\phi}&   & c_{\phi}
   \end{pmatrix}
   \begin{pmatrix}
     c_{\omega} & s_{\omega} & \\
     -s_{\omega}& c_{\omega} & \\
                &            & 1
   \end{pmatrix} U_{\mathrm{Majorana}},
\qquad (\alpha = e, \mu, \tau, \quad i = 1, 2, 3).
\end{align}
Here, $s$ and $c$ denote $\sin$ and $\cos$, and
$U_{\rm Majorana}$ is the so-called Majorana phase matrix, which does not
contribute to the oscillation phenomena.
The effective Hamiltonian for neutrino propagation is
\begin{align}
H(x)_{\beta \alpha} =&
\frac{1}{2 E}
     \left\{
       U_{\beta i}
        \begin{pmatrix}
            0 & & \\
              & \Delta m_{21}^{2} & \\
              &                   & \Delta m_{31}^{2}
        \end{pmatrix}
       U^{\dagger}_{i \alpha}
                +
        \begin{pmatrix}
         a_{0} + \delta a (x) &   & \\
                              & 0 & \\
                              &   & 0
        \end{pmatrix}_{\beta \alpha}
       \right\},
\end{align}
where $\Delta m_{ij}^{2}$ is the squared-mass difference between the $i$th
and $j$th generations in a vacuum.  We separate the matter effect into two
parts, $a_{0}$ and $\delta a(x)$.  The effect of the average matter
is denoted by $a_{0}$, and $\delta a(x)$ is the matter profile
part, that is, the deviation part from the average density which depends
on the position, $x$.  After the separation, we expand the matter
profile part into the Fourier series,
\begin{gather}
\delta a (x) = \sum_{\begin{minipage}{1.2cm} 
{\footnotesize $n = - \infty \vspace{-0.1cm}\\ 
\hspace*{0.2cm}n \ne 0 $}\end{minipage}}^{\infty} a_{n} e^{-i p_{n} x},
\qquad
p_{n} \equiv \frac{2 \pi}{L} n .
\end{gather}
Note that the relation, $a_{n} = a_{-n}^{*}$ is always satisfied because of
the condition that $\delta a(x)$ is real.
For anti-neutrino, $a_{0}$, $a_{n}$ and $\delta$ should be replaced by
$- a_{0}$, $- a_{n}$ and $- \delta$, respectively. 

We treat $\Delta m_{21}^{2}$ and $\delta a(x)$ as perturbations and
obtain the oscillation probabilities up to the first order of them,
for example \cite{OS},
\begin{align}
P_{\nu_{e} \rightarrow \nu_{\mu}}
 &= s_{\psi}^{2} s_{2 \tilde{\phi}}^{2}
   \sin^{2} \frac{\lambda_{+}-\lambda_{-}}{4E} L 
                                    & \equiv P^{\rm main}\nonumber \\
 & \qquad +
  \frac{1}{2} c_{\delta} s_{2 \psi} s_{2 \omega} s_{2 \tilde
   {\phi}} & \nonumber \\
 & \qquad \qquad \times \Biggl[
    \left( c_{\tilde{\phi}} c_{\phi - \tilde{\phi}}
    \frac{\Delta m_{21}^{2}}{\Delta m_{21}^{2} c_{\omega}^
     {2}-\lambda_{-}}
      +
      s_{\tilde{\phi}} s_{\phi - \tilde{\phi}}
    \frac{\Delta m_{21}^{2}}{\lambda_{+}-\Delta m_{21}^{2}
    c_{\omega}^{2}}
    \right)
    \sin^{2} \frac{\Delta m_{21}^{2} c_{\omega}^
     {2}-\lambda_{-}}{4E} L \nonumber & \\
 & \qquad \qquad \quad -
  \left( c_{\tilde{\phi}} c_{\phi - \tilde{\phi}}
    \frac{\Delta m_{21}^{2}}{\Delta m_{21}^{2} c_{\omega}^
     {2}-\lambda_{-}}
      +
      s_{\tilde{\phi}} s_{\phi - \tilde{\phi}}
    \frac{\Delta m_{21}^{2}}{\lambda_{+}-\Delta m_{21}^{2}
    c_{\omega}^{2}}
    \right)
    \sin^{2} \frac{\lambda_{+}-\Delta m_{21}^{2} c_{\omega}^
     {2}}{4E} L \nonumber & \\
 & \qquad \qquad \quad +
  \left( c_{\tilde{\phi}} c_{\phi - \tilde{\phi}}
    \frac{\Delta m_{21}^{2}}{\Delta m_{21}^{2} c_{\omega}^
     {2}-\lambda_{-}}
      -
      s_{\tilde{\phi}} s_{\phi - \tilde{\phi}}
    \frac{\Delta m_{21}^{2}}{\lambda_{+}-\Delta m_{21}^{2}
    c_{\omega}^{2}}
    \right)
    \sin^{2} \frac{\lambda_{+}-\lambda_{-}}{4E} L
    \Biggr] \nonumber  & \\
 & \qquad +\frac{1}{4}
  s_{\delta} s_{2 \psi} s_{2 \omega} s_{2 \tilde{\phi}}
  \left( c_{\tilde{\phi}} c_{\phi - \tilde{\phi}}
    \frac{\Delta m_{21}^{2}}{\Delta m_{21}^{2} c_{\omega}^
     {2}-\lambda_{-}}
  +
      s_{\tilde{\phi}} s_{\phi - \tilde{\phi}}
    \frac{\Delta m_{21}^{2}}{\lambda_{+}-\Delta m_{21}^{2}
    c_{\omega}^{2}}
    \right) \nonumber & \\
 & \qquad \qquad \times
  \left(
   \sin \frac{\lambda_{+}-\Delta m_{21}^{2} c_{\omega}^
     {2}}{2E} L
   + \sin \frac{\Delta m_{21}^{2} c_{\omega}^
     {2}-\lambda_{-}}{2E} L
   - \sin \frac{\lambda_{+}-\lambda_{-}}{2E} L
  \right) \nonumber & \\
 & \qquad +
  4 s_{\psi}^{2} s_{2 \tilde{\phi}}^{2} c_{2 \tilde{\phi}}
  \sum_{n=1}^{\infty} {\rm Re}[ a_{n} ]
   \frac{ \lambda_{+}-\lambda_{-} }
        {(\lambda_{+}-\lambda_{-})^{2} - (2E p_{n})^{2}}
  \sin^{2} \frac{\lambda_{+} - \lambda_{-}}{4E} L
   & \equiv \sum_{n = 1}^{\infty} P^{\rm profile}_{n},
\label{eq:oscProb-nue-to-numu}\\
P_{\nu_{e} \rightarrow \nu_{\tau}}
 &= c_{\psi}^{2} s_{2 \tilde{\phi}}^{2}
   \sin^{2} \frac{\lambda_{+}-\lambda_{-}}{4E} L & \nonumber \\
 & \qquad -
  \frac{1}{2} c_{\delta} s_{2 \psi} s_{2 \omega} s_{2 \tilde
   {\phi}} & \nonumber \\
 & \qquad \qquad \times \Biggl[
    \left( c_{\tilde{\phi}} c_{\phi - \tilde{\phi}}
    \frac{\Delta m_{21}^{2}}{\Delta m_{21}^{2} c_{\omega}^
     {2}-\lambda_{-}}
      +
      s_{\tilde{\phi}} s_{\phi - \tilde{\phi}}
    \frac{\Delta m_{21}^{2}}{\lambda_{+}-\Delta m_{21}^{2}
    c_{\omega}^{2}}
    \right)
    \sin^{2} \frac{\Delta m_{21}^{2} c_{\omega}^
     {2}-\lambda_{-}}{4E} L & \nonumber \\
 & \qquad \qquad \quad -
  \left( c_{\tilde{\phi}} c_{\phi - \tilde{\phi}}
    \frac{\Delta m_{21}^{2}}{\Delta m_{21}^{2} c_{\omega}^
     {2}-\lambda_{-}}
      +
      s_{\tilde{\phi}} s_{\phi - \tilde{\phi}}
    \frac{\Delta m_{21}^{2}}{\lambda_{+}-\Delta m_{21}^{2}
    c_{\omega}^{2}}
    \right)
    \sin^{2} \frac{\lambda_{+}-\Delta m_{21}^{2} c_{\omega}^
     {2}}{4E} L &\nonumber \\
 & \qquad \qquad \quad +
  \left( c_{\tilde{\phi}} c_{\phi - \tilde{\phi}}
    \frac{\Delta m_{21}^{2}}{\Delta m_{21}^{2} c_{\omega}^
     {2}-\lambda_{-}}
      -
      s_{\tilde{\phi}} s_{\phi - \tilde{\phi}}
    \frac{\Delta m_{21}^{2}}{\lambda_{+}-\Delta m_{21}^{2}
    c_{\omega}^{2}}
    \right)
    \sin^{2} \frac{\lambda_{+}-\lambda_{-}}{4E} L
    \Biggr] &\nonumber \\
 & \qquad -\frac{1}{4}
  s_{\delta} s_{2 \psi} s_{2 \omega} s_{2 \tilde{\phi}}
  \left( c_{\tilde{\phi}} c_{\phi - \tilde{\phi}}
    \frac{\Delta m_{21}^{2}}{\Delta m_{21}^{2} c_{\omega}^
     {2}-\lambda_{-}}
  +
      s_{\tilde{\phi}} s_{\phi - \tilde{\phi}}
    \frac{\Delta m_{21}^{2}}{\lambda_{+}-\Delta m_{21}^{2}
    c_{\omega}^{2}}
    \right) &\nonumber \\
 & \qquad \qquad \times
  \left(
   \sin \frac{\lambda_{+}-\Delta m_{21}^{2} c_{\omega}^
     {2}}{2E} L
   + \sin \frac{\Delta m_{21}^{2} c_{\omega}^
     {2}-\lambda_{-}}{2E} L
   - \sin \frac{\lambda_{+}-\lambda_{-}}{2E} L
  \right) &\nonumber \\
 & \qquad +
  4 c_{\psi}^{2} s_{2 \tilde{\phi}}^{2} c_{2 \tilde{\phi}}
  \sum_{n = 1}^{\infty} {\rm Re}[ a_{n} ]
   \frac{\lambda_{+}-\lambda_{-}}
        {(\lambda_{+}-\lambda_{-})^{2} - (2E p_{n})^{2}}
  \sin^{2} \frac{\lambda_{+} - \lambda_{-}}{4E} L, &
\label{eq:oscProb-nue-to-nutau}\\
P_{\nu_{e} \rightarrow \nu_{e}}
 &= 
  1 - s_{2 \tilde{\phi}}^{2} \sin^{2} 
       \frac{\lambda_{+} - \lambda_{-}}{4E} L & \nonumber \\
 & \qquad 
   - 4 s_{2 \tilde{\phi}}^{2} c_{2 \tilde{\phi}}
     \sum_{n = 1}^{\infty} {\rm Re}[a_{n}] 
       \frac{\lambda_{+} - \lambda_{-}}
            {(\lambda_{+} - \lambda_{-})^{2} - (2Ep_{n})^{2}}
        \sin^{2} \frac{\lambda_{+} - \lambda_{-}}{4E}L.
\label{eq:oscProb-nue-to-nue}
\end{align}
Here the effective mixing angle in the average density of the matter, 
$\tilde{\phi}$, 
and the squared-mass eigenvalue, $\lambda_{\pm}$, in the average
density are given by
\begin{gather}
\tan 2 \tilde{\phi} 
 = \frac{s_{2 \phi} ( \Delta m_{31}^{2} - \Delta
   m_{21}^{2} s_{\omega}^{2})}{c_{2 \phi} ( \Delta m_{31}^{2} - \Delta
   m_{21}^{2} s_{\omega}^{2}) - a_{0}}, \\
\lambda_{\pm} 
 = \frac{1}{2} \left[
      \Delta m_{31}^{2} + \Delta m_{21}^{2} s_{\omega}^{2} + a_{0}
        \pm
       \sqrt{ \left\{
            ( \Delta m_{31}^{2} - \Delta m_{21}^{2} s_{\omega}^
                {2} ) c_{2 \phi} - a_{0}
            \right\}^{2}
            +
            ( \Delta m_{31}^{2} - \Delta m_{21}^{2} s_{\omega}^
             {2} )^{2} s_{2 \phi}^{2}
           } \right].
\end{gather}
Note that the effects of the asymmetric profile do not appear in the first
order of the perturbations.

The resonance conditions, $\lambda_{+} - \lambda_{-} = 2 E p_{n}$, show
the energy range and width of the resonance induced by each Fourier
mode.  We understand that the higher Fourier mode can resonate only with
the lower energy neutrinos whose oscillation lengths are shorter
\cite{Smirnov}.  Furthermore, the half-width of the amplitude becomes
narrow as the mode becomes higher.
This means that if the lower energy neutrino were to be observed precisely,
the fine structure of the Earth could be known, although it is actually
extremely difficult to achieve such observation.  
Therefore, we conclude that only the first few Fourier modes,
which are determined by the large structure of the matter profile, are
relevant in the currently assumed experimental set-ups.  This is consistent
with the results obtained using different methods \cite{Shan, Jacobsson,
reconstruction}.
								   
\subsection{Uncertainty of the Earth model}
Knowledge of geophysics is essential since it is very difficult to ascertain
the profile of the Earth from a neutrino experiment.  So far, PREM
has been regarded as the absolute model.  We have tended to expect that
the error and the effect induced by the error are so small that they can be
neglected without a careful consideration, although we have to estimate how
much error the model includes in order to use the seismological Earth model.
In order to discuss this error, we introduce another Earth model, ak135-f
\cite{ak135}.

\begin{figure*}[th]
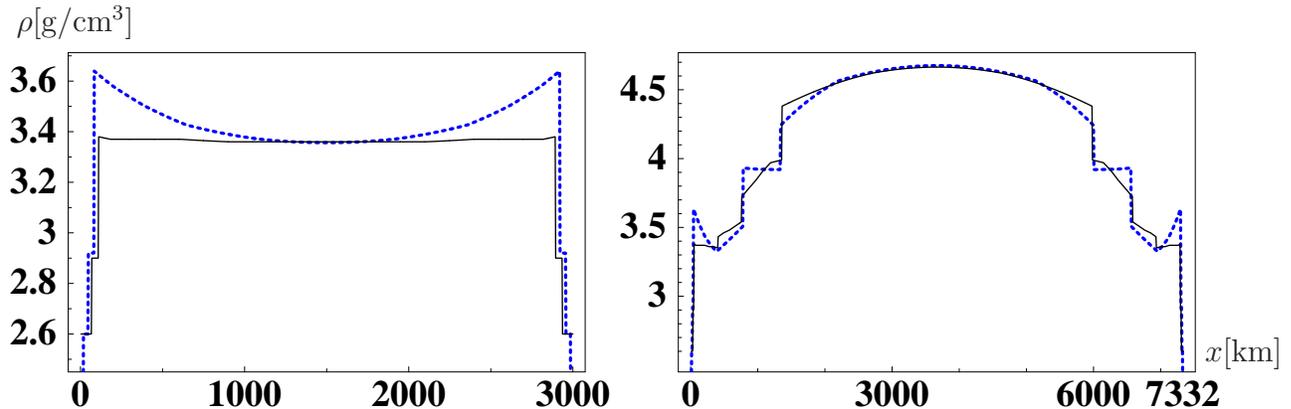

\unitlength=1cm
\begin{picture}(16,5)
\includegraphics[width=8cm]{profile-3000km-PREM-vs-ak135f.epsi}
\includegraphics[width=8cm]{profile-7332km-PREM-vs-ak135f.epsi}
\put(-0.2,0.6){\large $x$[km]}
\put(-16,5){\large $\rho {\rm [g/cm^{3}]}$}
\end{picture}
\caption{%
The matter profile functions in the case of 3,000 km and 7,332 km 
which are calculated using PREM (solid line) and ak135-f (dotted line).
}
 \label{Fig:profileFunction-PREM-vs-ak135f}
\end{figure*}
\begin{figure*}[th]
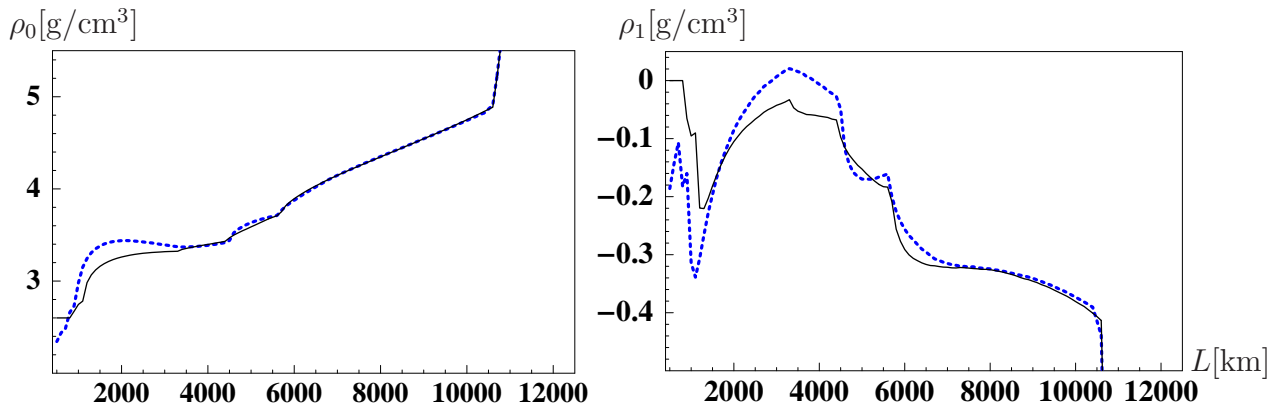

\unitlength=1cm
\begin{picture}(16,5)
\hspace{0.25cm}
\includegraphics[width=7.5cm]{averageDens-PREM-vs-ak135f.epsi}
\includegraphics[width=8cm]{firstFourierCoe-PREM-vs-ak135f.epsi}
\put(-0.1,0.4){\large $L$[km]}
\put(-15.8,4.9){\large $\rho_{0} {\rm [g/cm^{3}]}$}
\put(-7.7,4.9){\large $\rho_{1} {\rm [g/cm^{3}]}$}
\end{picture}
\caption{%
 The baseline dependence of the average matter density and the
first Fourier coefficient which are calculated using PREM (solid line) 
and ak135-f (dotted line).
 Around $L=3,000$ km, the average densities of these two models are 
almost the same within a few percent, but the first Fourier coefficients 
differ by more than 100\%.
}
 \label{Fig:averageDen-and-firstFourieCoe-PREM-vs-ak135f}
\end{figure*}
\begin{figure*}[th]
\unitlength=1cm
\begin{picture}(8,5)
\hspace{0.25cm}
\includegraphics[width=8cm]{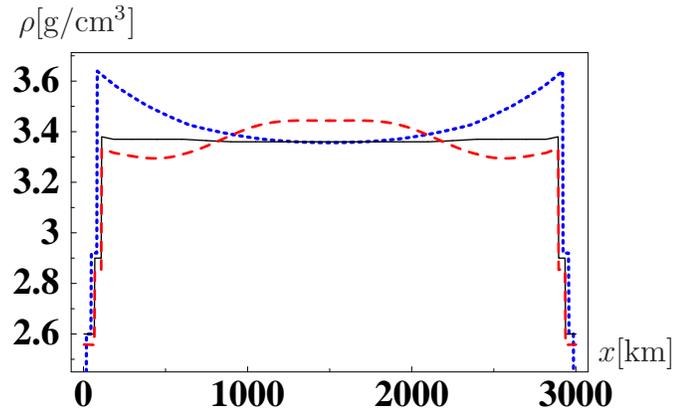}
\put(-0.2,0.6){\large $x$[km]}
\put(-7.9,5){\large $\rho {\rm [g/cm^{3}]}$}
\end{picture}
\caption{%
An example where the deviation of the first Fourier coefficient from PREM 
is more than 100\% in $L = $ 3,000 km (dashed line). This is drawn
by modifying PREM within 2.5\%.
The first Fourier coefficient for this profile is $-0.084$ ${\rm g/cm^{3}}$,
which is almost twice as large as that of PREM, $-0.043$ ${\rm g/cm^{3}}$.
Solid and dotted lines are the same as those in 
Fig.\ref{Fig:profileFunction-PREM-vs-ak135f}.
}
 \label{Fig:profileFunction-PREM-vs-ak135f-vs-Example}
\end{figure*}

Figure \ref{Fig:profileFunction-PREM-vs-ak135f} represents the matter
profile function calculated using PREM (solid line) and ak135-f (dotted
line) in the case where the baseline length is 3,000 km and 7,332 km,
respectively.  The profile function based on PREM is almost flat at $L=$
3,000 km, and this fact guarantees that the matter profile effect is
small. According to ak135-f, however, it is not so flat.  If we follow
this model, we may be unable to ignore the matter profile effect,
even at this baseline length.  Furthermore, the authors of
Ref.\cite{ak135} note that ``the upper mantle density model should be
treated with caution and may well change with further work.''  The upper
mantle and transition area (up to 670 km in depth) occupy a large part
of the path of the neutrino beam in the case of $L=$ 3,000 km.

The baseline dependence of the average density and the first Fourier
coefficient are compared 
in Fig.\ref{Fig:averageDen-and-firstFourieCoe-PREM-vs-ak135f}.  These two
models are different in terms of the Fourier coefficient, although they
are similar with regard to the average matter density.
In particular, around $L=$ 3,000 km, the difference of the first Fourier
coefficient is quite large.  We need to recognize that seismological
models include uncertainty, the size of which is at least equal to the
difference in these two models%
\footnote{
Assumption for the profile of the underground chemical component is
made when the density profile is determined in geophysics.  This
assumption affects the density of the electron number that we really
want to ascertain.}.  The authors of Ref.\cite{Geller} state that $\pm 2 \sim
3$\% uncertainty in the upper mantle is a reasonable estimation, which
produces a large, as much as 100\%, uncertainty for the first Fourier
coefficient.  If 2.5\% error for PREM is allowed, we can assume the
matter profile depicted
Fig.\ref{Fig:profileFunction-PREM-vs-ak135f-vs-Example} instead of PREM
itself.  This profile realizes 100\% shift of the first Fourier
coefficient without adjusting the average matter density.  
{\it We would like
to stress again that the uncertainty of the average density may indeed
be small, but the uncertainty of the profile is not so small that it can be
neglected, especially in $L \simeq $3,000 km.}

\section{Correlation between average matter parameter and first Fourier
 coefficient} \label{Sec:Colleration}

Here, we will establish the existence of the correlation between the
parameter for average matter density and the first Fourier coefficient
of the matter profile function.  The effect induced by the matter
profile and its uncertainty can be understood by the concept of this
correlation.  {\it Although the average density can be determined at a
few percent accuracy through studies of the seismic wave and
gravitational effect etc., the matter profile effect adds extra
uncertainty to the average matter density.}  As we pointed out in the
previous section, the uncertainty of the first Fourier mode is not
small, especially in the case where the neutrino beam passes mainly through
the upper mantle and the transition zone, namely where the baseline length
is around 3,000 km.
 
\subsection{One example: $L=$ 7,332 km}
\begin{figure*}[th]
\unitlength=1cm
\begin{picture}(12,7)
\includegraphics[width=12cm]{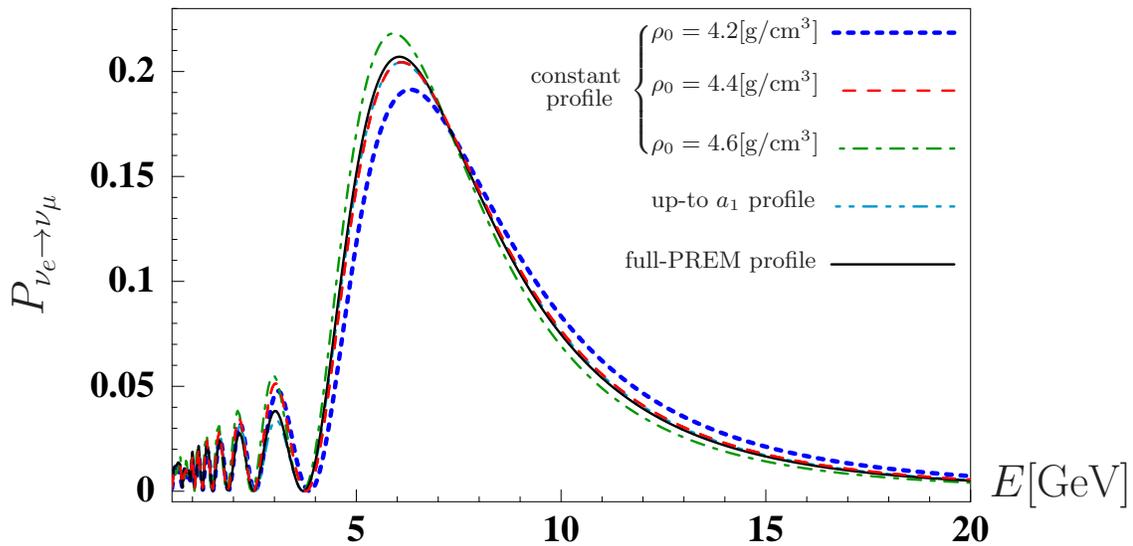}
\put(0,0.6){\LARGE $E$[GeV]}
\put(-13,3){\rotatebox{90}{\LARGE $P_{\nu_{e} \rightarrow \nu_{\mu}}$}}
\put(-6.1,6.1){constant}
\put(-5.9,5.8){profile}
\put(-5,6.1){\begin{minipage}{1cm}
            \begin{align}
            \left\{
              \begin{array}{c}
                     \\
                     \\
                     \\
                     \\
                     \\
              \end{array}
            \right. \nonumber 
            \end{align}
           \end{minipage}
           }
\put(-4.5,6.7){$\rho_{0} = 4.2{\rm[g/cm^{3}]}$}
\put(-4.5,6){$\rho_{0} = 4.4{\rm[g/cm^{3}]}$}
\put(-4.5,5.2){$\rho_{0} = 4.6{\rm[g/cm^{3}]}$}
\put(-4.5,4.45){up-to $a_{1}$ profile}
\put(-4.85,3.65){full-PREM profile}
\end{picture}
\caption{%
Oscillation probability for $\nu_{e} \rightarrow \nu_{\mu}$ in $L=7,332$
km.
The dotted, dashed, and dash-dot lines are calculated using the constant
 matter profile where the values of the matter parameter are 4.2, 4.4,
 and 4.6 $\rm{g/cm^3}$, respectively.  
The dash-dot-dot line is calculated by the up-to-first-mode profile
 where the values are $\rho_0=4.2$ $\rm{g/cm^3}$ and $\rho_1=-0.32$ $\rm{g/cm^3}$.
The solid line uses the full-PREM matter profile, and this line is
quite similar to the dash-dot-dot and dashed lines.
We set
the oscillation parameters as $\sin \omega = 0.5$, $\sin \psi =1/\sqrt{2}$,
 $\sin \phi = 0.1$, $\Delta m_{31}^{2} = 2.5 \times 10^{-3}{\rm eV}^{2}$, 
$\Delta m_{21}^{2} = 5.0 \times 10^{-5} {\rm eV}^{2}$,
and $\delta = \pi / 2$.}
\label{Fig:oscProb-PREM-vs-a0}
\end{figure*}

To corroborate that the correlation exists, we first present an
example.  In Fig.\ref{Fig:oscProb-PREM-vs-a0}, the oscillation
probability for $\nu_{e} \rightarrow \nu_{\mu}$ is shown where the baseline
length is 7,332 km.  The solid line is calculated using the full-PREM
profile.  As we showed in Ref.\cite{OS}, this line is almost the same as
the dash-dot-dot line which is calculated by the up-to-first-mode profile
with $\rho_{0}=4.2$ $\rm{g/cm^{3}}$ and $\rho_{1}=-0.32$
$\rm{g/cm^{3}}$, whose values are based on PREM.

The dotted, dashed, and dash-dot lines are those which are calculated 
using the constant-matter profile where the average density is $4.2$,
$4.4$, and $4.6$ $\rm{g/cm^{3}}$, respectively.  The PREM tells us that the
average density in this baseline length is 4.2 $\rm{g/cm^{3}}$.  This
figure shows that the constant profile with the average density
following PREM (dotted line) cannot reproduce the behavior of the
oscillation probability with the full-PREM profile (solid line).  However,
Fig.\ref{Fig:oscProb-PREM-vs-a0} also shows that if a shift in the
constant value is allowed, then a good fit with the full-PREM
calculation can be realized at $\rho_{0} = 4.4$ $\rm{g/cm^{3}}$.  This
fact means that the shift of the constant matter parameter can copy
the matter profile effect, which is almost equal to the effect induced
by the first Fourier modes with this baseline length.  We can expect that
there is a strong correlation between the parameter for the average
matter density and the first Fourier coefficient of the matter profile
function.  Indeed, this correlation is not an accidental phenomenon in
this example. 

We now consider the mechanism on which this correlation is
based.  The existence of the correlation suggests that the common
shift in the average matter parameter over a wide energy region can
imitate the effect of the first Fourier mode.  This statement means
that the relation,
\begin{align}
\frac{\partial P(a_{n} = 0)}{\partial a_{0}} \Delta a_{0}
 = \sum_{n = 1}^{\infty} P^{\rm profile}_{n}, \label{eq:correlationCondition}
\end{align}
is satisfied by the constant $\Delta a_{0}$ over a wide energy region.
In the above example, the constant shift $\Delta \rho_{0} = 0.2$ ${\rm
g/cm^{3}}$ works well within the energy region above 4 GeV. 
 Since we can assume that the unperturbed term of the oscillation
probability is dominant over the other perturbative terms and that the matter
 profile effect can be represented by the first mode, the condition in 
eq.\eqref{eq:correlationCondition} reduces to
\begin{align}
\frac{\partial P^{\rm main}}{\partial a_{0}} \Delta a_{0}
   &= P^{\rm profile}_{1} \nonumber\\
\Longleftrightarrow \quad 
  \frac{\Delta a_{0}}{\lambda_{+} - \lambda_{-}}
    \sin \frac{\lambda_{+} - \lambda_{-}}{4 E} L
   -
   \left(
    \frac{\Delta a_{0}}{4E}L
   \right) 
     \cos \frac{\lambda_{+} - \lambda_{-}}{4 E} L
 &=
   2 \left( \frac{{\rm Re}[a_{1}]}{4E} L \right)
     \frac{\frac{\lambda_{+} - \lambda_{-}}{4 E} L}
          {\left(
             \frac{\lambda_{+} - \lambda_{-}}{4 E} L
           \right)^{2}
              - 
           \pi^{2}}
     \sin \frac{\lambda_{+} - \lambda_{-}}{4 E} L. 
 \label{eq:reducedCorrelationCondition}
\end{align}
It is not inconsequential whether or not the constant shift 
$\Delta a_{0}$ can maintain the relation. 
To clarify this issue, we divide the energy region into two regions
where $\frac{\lambda_{+}-\lambda_{-}}{4E}L \sim 0$ and
$\frac{\lambda_{+}-\lambda_{-}}{4E}L \sim \frac{\pi}{2}$ are satisfied,
and we investigate each region.
\begin{figure*}[th]
\unitlength=1cm
\begin{picture}(12,7)
\hspace{-0.3cm}
\includegraphics[width=12.2cm]{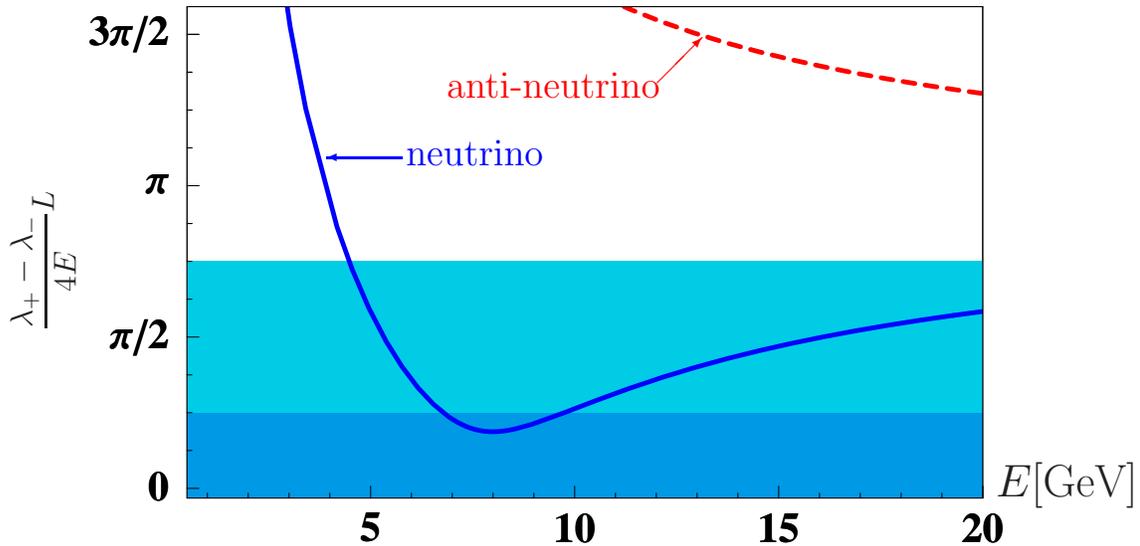}
\put(-0.1,0.6){\LARGE $E$[GeV]}
\put(-13.2,2.8){\rotatebox{90}{\large $\displaystyle \frac{\lambda_{+} - \lambda_{-}}{4E} L$}}
\put(-7.94,5){\Large \color[rgb]{0,0,1}neutrino}
\put(-8,5.1){\color[rgb]{0,0,1}\vector(-1,0){1}}
\put(-7.4,5.9){\Large \color[rgb]{1,0,0}anti-neutrino}
\put(-4.6,6.1){\color[rgb]{1,0,0} \vector(1,1){0.6}}
\end{picture}
\caption{%
Energy dependence of $\frac{\lambda_{+}-\lambda_{-}}{4E}L $ for neutrino
(solid line) and anti-neutrino (dashed line) in $L=$ 7,332 km.  The
oscillation parameters used in this figure are the same as those used in
Fig.\ref{Fig:oscProb-PREM-vs-a0}.
}
\label{Fig:energyDep-of-phase-7332}
\end{figure*}

Figure \ref{Fig:energyDep-of-phase-7332} shows the energy dependence of
$\frac{\lambda_{+}-\lambda_{-}}{4E}L$ corresponding to the case of
Fig.\ref{Fig:oscProb-PREM-vs-a0}.  In the region of
$\frac{\lambda_{+}-\lambda_{-}}{4E}L \sim 0$, the condition
eq.\eqref{eq:reducedCorrelationCondition} is approximated to
\begin{align}
 \Delta a_{0} = - \frac{6}{\pi^{2}} {\rm Re}[a_{1}] 
                + \mathcal{O} \left\{\left(
                                \frac{\lambda_{+} - \lambda_{-}}
                                     {4E} L
                              \right)^{2} \right\}.
\label{eq:Correlation0}
\end{align}
Using the value $\rho_{1} = - 0.32$ ${\rm g/cm^3}$, we understand 
from eq.\eqref{eq:Correlation0} that
shifting $\rho_0$ by $0.2{\rm g/cm^3}$ makes the oscillation probability
calculated with the constant profile clearly mimic the real oscillation
probability, including the matter profile.  Within the 
$\frac{\lambda_{+}-\lambda_{-}}{4E}L \sim \frac{\pi}{2}$ region, the
condition eq.\eqref{eq:reducedCorrelationCondition} is approximated by
\begin{align}
 \Delta a_{0} = - \frac{2}{3} {\rm Re}[a_{1}]
                              + \mathcal{O} \left(
                                \frac{\lambda_{+} - \lambda_{-}}
                                     {4E} L 
                                 -
                                 \frac{\pi}{2}
                              \right). 
\label{eq:CorrelationPi2}
\end{align}
This means that the shift of $a_{0}$ so as to mimic the matter profile
effect is also about $0.2$ ${\rm g/cm^3}$, which is the same as the
shift in the $\frac{\lambda_{+}-\lambda_{-}}{4E}L \sim 0$ region.
Therefore, in a wide energy region, such as both
the $\frac{\lambda_{+}-\lambda_{-}}{4E}L \sim 0$ and
$\frac{\lambda_{+}-\lambda_{-}}{4E}L \sim \frac{\pi}{2}$ regions, the
common shift of $a_{0}$ can indeed copy the effect of the first
Fourier mode\footnote{Moreover, we can see that this common shift works
well in 
the intermediate region by expanding the oscillation probability around
$\frac{\lambda_{+}-\lambda_{-}}{4E}L \sim \frac{\pi}{4}$.}.

It is an essential point for the existence of the correlation that an
experiment is only sensitive to these two energy ranges%
\footnote{In the region where
$\frac{\lambda_{+}-\lambda_{-}}{4E}L \sim \pi $ holds, the relation
becomes $\Delta a_{0} = - \frac{\pi}{\pi + 1} {\rm Re}[a_{1}] 
                             + \mathcal{O}
                                \left(
                                \frac{\lambda_{+} - \lambda_{-}}
                                     {4E} L 
                                 -
                                 \pi
                                 \right)$, 
and hence the required shift to mimic the matter profile effect is
significantly different.}.
We note that the condition for anti-neutrino is the same as that for
neutrino.  Therefore, even if the analysis is made using both neutrino and
anti-neutrino, the correlation still exists if the observed
energy region for both neutrino and anti-neutrino satisfies either
$\frac{\lambda_{+}-\lambda_{-}}{4E}L \sim 0$ or 
$\frac{\lambda_{+}-\lambda_{-}}{4E}L \sim \frac{\pi}{2}$. 
In the case of $L=7,332$ km, the energy range for anti-neutrino is not
around either $\frac{\lambda_{+}-\lambda_{-}}{4E}L \sim 0$ or 
$\frac{\lambda_{+}-\lambda_{-}}{4E}L \sim \frac{\pi}{2}$, so if we could
observe an adequate number of anti-neutrino events, the correlation
would cease to hold, although the anti-neutrino event can be expected to 
be too short to be significant in a statistical sense. 
Therefore, we can conclude that the
correlation still exists for this baseline length.

\subsection{Baseline region where the correlation exists}
There is a strong correlation between
$a_{0}$ and $a_{1}$ within the energy and baseline region where
$\frac{\lambda_{+}- \lambda_{-}}{4E}L \sim 0$, $\frac{\pi}{2}$ 
are satisfied, as we have established in the previous sub-section.
\begin{figure*}[th]
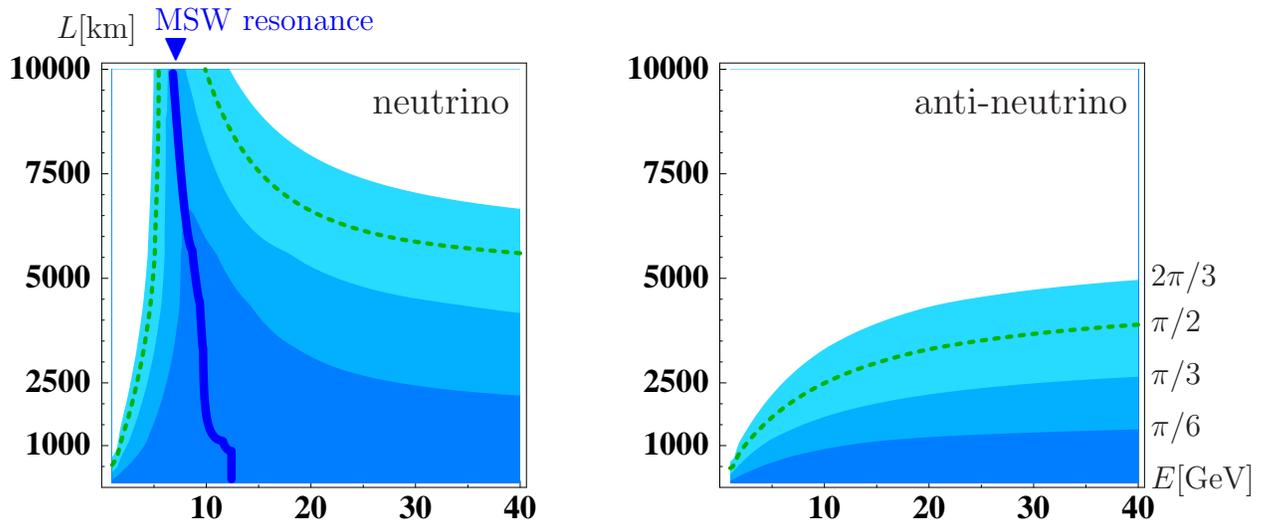

\unitlength=1cm
\begin{picture}(15,7)
\includegraphics[width=7cm]{correlation-region-nu-dm31-25e-4.epsi}
\hspace{1cm}
\includegraphics[width=7cm]{correlation-region-antinu-dm31-25e-4.epsi}
\put(-0.05,0.4){\large $E$[GeV]}
\put(-14.6,6.4){\large $L$[km]}
\put(-0.05,1.1){\large $\pi/6$}
\put(-0.05,1.8){\large $\pi/3$}
\put(-0.05,2.5){\large $\pi/2$}
\put(-0.05,3.1){\large $2 \pi/3$}
\put(-13.2,6.1){\color[rgb]{0,0,1}{\Large$\blacktriangledown$}}
\put(-13.3,6.5){\color[rgb]{0,0,1}{\large MSW resonance}}
\put(-10.4,5.4){\Large neutrino}
\put(-3.2,5.4){\Large anti-neutrino}
\end{picture}
\caption{%
Energy and baseline dependence of $\frac{\lambda_{+}- \lambda_{-}}{4E}L$
for neutrino (left plot) and anti-neutrino (right plot).  The reference
values of the parameters are the same as in
Fig.\ref{Fig:oscProb-PREM-vs-a0}.
}
 \label{Fig:energy-baseline-dep-of-phase}
\end{figure*}
The energy and baseline dependence of $ \frac{\lambda_{+}-
\lambda_{-}}{4E}L $ for neutrino and anti-neutrino is shown in 
Fig.\ref{Fig:energy-baseline-dep-of-phase}.  This figure tells us the region
where the correlation exists.  In $L \lesssim $ 5,000 km, $
\frac{\lambda_{+}- \lambda_{-}}{4E}L $ is near 0 or $\frac{\pi}{2}$ for
both neutrino and anti-neutrino throughout almost all the energy range.  
Therefore,
the common shift of $a_{0}$ for neutrino and anti-neutrino can mimic the
effect of $a_{1}$, that is, the correlation does exist. In 5,000 km
$\lesssim L \lesssim$ 7,500 km, $ \frac{\lambda_{+}- \lambda_{-}}{4E}L $
for anti-neutrino is not around either 0 or $\frac{\pi}{2}$.  However, 
since there will not be a significant number of anti-neutrino events, in this
region, the statistics will be dominated by neutrino events and hence the
correlation still exists.  In the region beyond 7,500 km, the high-energy
neutrino no longer follows the condition, and so the correlation will
cease to hold.
{\it We conclude that there is a strong correlation between $a_{0}$ and
$a_{1}$ in the case where the baseline length is less than 7,500 km.}

We would like to note one more thing.  According to
eqs.\eqref{eq:oscProb-nue-to-nutau} and \eqref{eq:oscProb-nue-to-nue},
the same relation also holds for $\nu_{e} \rightarrow \nu_{\tau} $ and
$\nu_{e} \rightarrow \nu_{e} $.  So, even if we utilize these channels,
the correlation will not be broken.

\section{Numerical calculations} \label{Sec:Numerical}
To illustrate how the uncertainty of $a_{1}$ impairs sensitivity to
the CP violation, we present the numerical results for it, including the
matter profile effect and its uncertainty.  In an effort to clarify 
the difference from former research, we show simultaneously the result 
without consideration of the matter profile.

First, we explain briefly the procedure for drawing the sensitivity plot:
We define the test statistics,
\begin{align}
\chi^{2} &\equiv
          \min_{\begin{minipage}{2cm} 
                \begin{center}
                {\footnotesize osc. param.}\\
                {\footnotesize $\delta = \{0, \pi\}$}
                \end{center}  
               \end{minipage}}
        \left[ 
         \sum_{i}^{\rm bin}
         \frac{\left| \bar{N}^{{\rm th}}_{i} \times N^{\rm ex}_{i}
                - N^{{\rm th}}_{i} \times \bar{N}^{\rm ex}_{i}
               \right|^{2}}
              { (\bar{N}^{{\rm th}}_{i} )^2 \times N^{\rm ex}_i
              +(N^{{\rm th}}_{i})^2 \times \bar{N}^{\rm ex}_i}
       \right].
\label{eq:chi3}
\end{align}
Here, $N^{\rm ex}$ is the ``expected number of events'' for 
$\nu_{e} \rightarrow \nu_{\mu}$ calculated using the full-PREM 
matter profile with $\delta = \pi / 2$, $N^{\rm th}$ is the 
``theoretical number of events'',
calculated with the constant matter profile or the up-to-first-mode
profile with $\delta=\{0, \pi \}$, and $\bar{N}$ denotes the number of
events for
$\bar{\nu}_{e} \rightarrow \bar{\nu}_{\mu}$.  The index $i$ stands for
the energy bin.  The parameters contained in $ N^{\rm th} $ and
$\bar{N}^{\rm th}$ are varied within given ambiguities.  We adjust them
and minimize $\chi^{2}$ to introduce the effect of the parameter
correlation \cite{nuFact, KOS}.
 The widths of uncertainty of the parameters concerning the atmospheric
and solar neutrino experiments are expected to be narrowed by near-future 
experiments \cite{near-future}.
Therefore, we assume
\begin{gather}
 \Delta (\sin \psi) = 1 \%, \quad 
 \Delta (\Delta m_{31}^{2}) = 3 \%, \nonumber\\
 \Delta (\sin \omega) = 5 \%, \quad
 \Delta (\Delta m_{21}^{2}) = 5 \% ,
\label{eq:width-of-uncertainty-No1}
\end{gather}
and for the other parameters including the matter effect we assume some
options and compare them to each other.
We do not deal with systematic error.
We require 
\begin{align}
 \chi^{2} > \chi^{2}_{99\%}({\rm d.o.f = bin})
\label{eq:condition-for-CPobserve}
\end{align}
in order to claim that the hypothesis with $\delta = \{0, \pi\}$ is
excluded at the $99\%$ level of significance, 
where the right-hand side is the $\chi^{2}$
distribution function whose degree of freedom is {\it the number of 
energy bins, not the number of parameters which fit.}  This is because
we adopt the concept, {\it the power of test} (see appendix of
Ref.\cite{KOS} for more details).  The reason why we use this, less
familiar, concept is that we firmly believe that we must pay attention to
the fact that the best-fit point suggested by an experiment is not always located
on the point chosen by nature. 
Remind yourself how the best-fit point
for the solar neutrino deficit has changed. 
We actually know that there were several good-fitting regions for the solar 
neutrino experiments which were separated from each other on the parameter 
plane and which were not distributed only around the best-fit point. 
Moreover, the best-fit point itself moved from one region to another. 
To discuss the feasibility of observing some quantity with an experiment, 
we have to consider this fact.

We also note in passing that the degree of freedom for so-called $\Delta
\chi^{2}$, which is often used in an estimation of the 
oscillation parameters, is not the number of the parameters 
since, for example, two strongly correlated parameters
are not independent of each other, and hence 
we should count them as one parameter. 
Indeed, as is commonly known, 
there are strong correlations for some parameters 
in the case we deal with\footnote{
Depending on the conditions, it breaks the correlation to regard the
oscillation parameters as not free ones, but as restricted ones.
In the numerical calculation, this effect is automatically introduced 
by setting the widths of the uncertainty.
}.
We should be more careful about what is a truly measurable quantity
\cite{JS}, and which quantities statistics are sensitive to \cite{KOS}.
For instance, if the statistics can be constructed so as to be sensitive 
to the CP violating effect, then the parameter is only one, Jarlskog's
parameter, and the degree of freedom should be one.  Strictly
speaking, only when the statistics are linearly dependent on the parameters,
will the degree of freedom coincide with the number of the parameters.

\begin{figure*}[th]
\unitlength=1cm
\hspace*{-2cm}
\begin{picture}(10,6.2)
 \put(-0.2,5.9){\Large$\Delta m_{21}^{2}{\rm [eV^{2}]}$}
 \put(10.2,0.2){\Large $\sin \phi (= |U_{e3}|)$}
 \includegraphics[width=10cm]{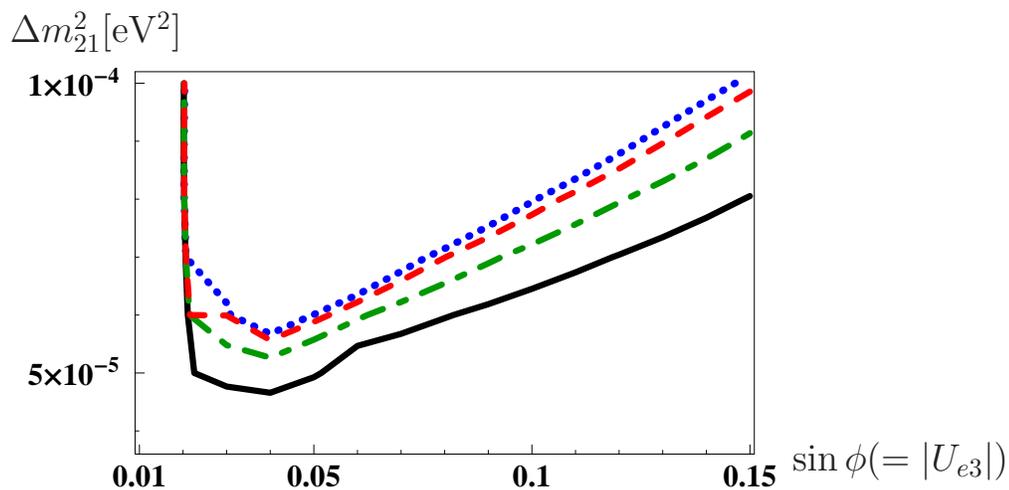}
\end{picture}
\caption{%
Sensitivity reach for CP violation with $\delta = \pi/2$,
a baseline length of 3,000 km, and a muon energy of 30 GeV.
The theoretical number of events, $N^{\rm th}$, of the dotted curve 
is calculated using the up-to-first-mode profile, where
the widths of parameter uncertainty are assumed to be $\Delta (a_{0}) = 3\%$, 
$\Delta (a_{1}) = 200\%$, and $\sin \phi = 0 \sim 0.16$.
For the other parameters, the uncertainty in
eq.\eqref{eq:width-of-uncertainty-No1} is assumed.
The solid, dash-dot and dashed curves are calculated using the
constant matter profile whose uncertainties are $\Delta a_{0} = 0\%, 3\%$
and $5\%$, respectively, and the uncertainty for the other parameters is 
the same as for the dotted line.
}
 \label{Fig:CPsensitivityPlot}
\end{figure*}
To draw a sensitivity plot, we assume 40 kt detector and $10^{21}$
muon decays in the neutrino factory scheme, and interpret
eq.\eqref{eq:condition-for-CPobserve} to the condition for $\Delta
m_{21}^{2}$ and $\sin \phi$.  Figure \ref{Fig:CPsensitivityPlot}
indicates the lower boundary of $\Delta m_{21}^{2}$ and $|U_{e3}|$ $(=
\sin \phi)$ to reject the hypotheses that $\delta$ is $0$ and $\pi$ at
a 99\% level of significance, when nature adopts the value 
$\delta = \pi/2$ in the
case where the baseline length is 3,000 km, muon energy is 30 GeV,
the detection threshold is 5 GeV and the width of energy bin is 2.5 GeV (10
bin).  The reference values are the same as those in
Fig.\ref{Fig:oscProb-PREM-vs-a0}, except for $\Delta m_{21}^{2}$ and
$\sin \phi$;
\begin{gather}
 \sin \psi = 1/\sqrt{2}, \quad 
 \Delta m_{31}^{2} = 2.5 \times 10^{-3} {\rm eV^{2}},\quad 
  \sin \omega = 0.5, \quad \delta = \pi/2.
\end{gather}

The dotted curve is calculated whilst considering the matter profile effect and
its uncertainty up to the first Fourier mode in the calculation of
$N^{\rm th}$.  We take 3\% for $a_{0}$ and 200\% for $a_{1}$ as the
widths of uncertainty. The other curves are calculated assuming the
constant matter profile with a different uncertainty for $a_{0}$.  
The solid, dash-dot, and dashed curves correspond to $\Delta (a_{0}) =$
0\%, 3\% and 5\%, respectively.  In the minimization process, $\sin \phi$ in
$N^{\rm th}$ is taken as an arbitrary value between $0 \sim 0.16$, 
and the uncertainty
of the other parameters is assumed, as in
eq.\eqref{eq:width-of-uncertainty-No1}.

The dotted curve differs from the dash-dot curve in which the same
uncertainty for $a_{0}$ is assumed, but that for $a_1$ is not included. 
In contrast, it is very similar to the dashed curve whose uncertainty
for $a_{0}$ is larger (5\%), but where a constant matter profile is assumed.
This fact can be explained by the correlation between $a_0$ and $a_1$
eqs.\eqref{eq:Correlation0} and \eqref{eq:CorrelationPi2}: {\it
Uncertainty of $a_{1}$ is translated into that of $a_{0}$ through the
correlation, and this extra uncertainty gives rise to an extra
absorption of the signal of the CP violating effect.}  We know that the
effect of the fake CP signal induced by the matter effect becomes more
serious as $|U_{e3}|$ increases, so if $|U_{e3}|$ is measured just
below the current boundary, we will have to take the effect induced by 
the density profile of the Earth more seriously.

In the case of $L=$ 3,000 km, the determination of $\sin \phi$ and the
reduction of its uncertainty do not contribute towards improving the
sensitivity much, since
the matter effect itself is large at this baseline length.  Furthermore,
lowering the detection threshold also hardly helps to improve the
sensitivity because the number of events in the high-energy region 
is overwhelmingly greater
than that in the low-energy region.  If we can realize a lower detection
threshold, an advantage can be gained with a shorter baseline length
and a lower energy beam.

\begin{figure*}[th]
\unitlength=1cm
\hspace*{-2cm}
\begin{picture}(10,6.2)
 \put(-0.2,5.9){\Large$\Delta m_{21}^{2}{\rm [eV^{2}]}$}
 \put(10.2,0.2){\Large $\sin \phi (= |U_{e3}|)$}
 \includegraphics[width=10cm]{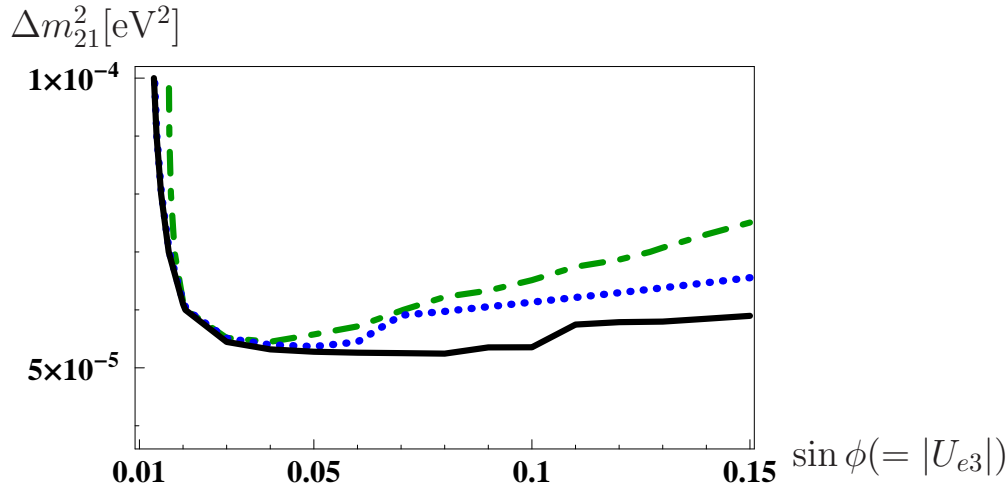}
\end{picture}
\caption{%
Sensitivity reach for CP violation when $\delta = \pi/2$ in the case
where the baseline length is 1,000 km, the muon energy is 11 GeV, and the
detection threshold is 1 GeV.  The up-to-first-mode profile is adopted
for the dotted curve where its widths of uncertainty are $\Delta (a_{0})
= 3\%$, $\Delta (a_{1}) = 200\%$, $\Delta(\sin \phi) = 10\%$ and
eq.\eqref{eq:width-of-uncertainty-No1} for the other parameters. 
The solid curve is calculated using the constant matter profile with $\Delta
(a_{0}) = 3\%$. By comparison with Fig.\ref{Fig:CPsensitivityPlot}, it
is obvious that the uncertainty of the matter effect is not serious in this
situation.  The dash-dot curve is calculated in a similar manner to the
dotted curve,
except that $\sin \phi$ varies from 0 to 0.16.  The dotted and dash-dot curves
show that the uncertainty of $\sin \phi$ will play an important role in this
situation if a large $\sin \phi$ is established.
}
 \label{Fig:CPsensitivityPlot-1000km}
\end{figure*}

Figure \ref{Fig:CPsensitivityPlot-1000km} shows the sensitivity in the
case where the baseline length is 1,000 km, the muon energy is 11 GeV and
the energy threshold is 1 GeV.
In this plot, the dotted curve represents the calculation using the
up-to-first-mode profile and its widths of parameter uncertainty are 
$\Delta (a_{0}) = 3\%$, $\Delta (a_{1}) = 200\%$, $\Delta (\sin \phi) =
10\%$, and eq.\eqref{eq:width-of-uncertainty-No1} for the others,
the solid curve is calculated by assuming a constant matter profile with
$\Delta (a_{0}) = 3\%$, and the dash-dot curve is the same as the dotted
one, except that $\sin \phi$ in $N^{\rm th}$ is assumed to be an
arbitrary value between $0 \sim 0.16$. 

Within the large $|U_{e3}|$ region, the sensitivity is better than that in
Fig.\ref{Fig:CPsensitivityPlot}.  Furthermore, the sensitivity of
this choice is more robust against the uncertainty of the matter effect 
than in the case of $L=$ 3,000 km. In contrast, 
although the signal of the CP violation 
is also small in the small $|U_{e3}|$ region, 
the fake CP violation effect induced by the matter effect 
is suppressed more strongly than the signal of the genuine CP violation, 
and hence, adopting a longer baseline and higher energy option may be
advantageous to the CP violation search.

This behavior does not come from the property of the statistics as
defined by eq.\eqref{eq:chi3}.  When the uncertainty of the constant
matter parameter is assumed to be larger, the optimization, using
so-called
$\Delta \chi^{2}$, also suggests that a shorter baseline length
and lower energy is better (see, for example O. Yasuda, in Ref\cite{nuFact}). 
The existence of this correlation tells us that 
even though it is said that the uncertainty of the average matter
density on the baseline is well estimated,
this is not the entire uncertainty of the constant matter parameter.
We would like to stress that the uncertainty of the matter effect is no
longer so small when we introduce the matter profile effect. 
Therefore, we should regard it
more seriously whatever statistics we use, especially in the case
where the main part of the neutrino beam path is the upper mantle and
transition zone, including a large uncertainty for the density profile.
We need to deal with the matter effect much more cautiously.


\section{Summary and conclusion} \label{Sec:Summary}
We pointed out that there was a very strong correlation between 
the constant matter parameter, $a_{0}$, and the first Fourier
coefficient of the matter profile function, $a_{1}$, 
within a wide energy and baseline region.
This fact means that the uncertainty of $a_{1}$ is translated into that of 
$a_{0}$, and this gives added uncertainty to $a_{0}$. 
 
We also showed that there is a huge uncertainty in $a_1$.  This is due to
the fact that seismological Earth models include a large uncertainty for
the density profile in the region, the upper mantle and the transition
zone, which is $24 \sim 670$ km in depth, and this uncertainty can cause
a huge, even a few hundred percent, uncertainty for $a_{1}$.  The existence
of the correlation suggests that this huge uncertainty affects CP
sensitivity since, due to the correlation, the uncertainty in $a_1$ gives
added uncertainty to $a_{0}$. 


We present the sensitivity plot for the CP violation effect including the
matter profile effect and its uncertainty.  In the case where the
baseline length is 3,000 km, most of the path of the neutrino beam 
is occupied by the upper mantle, which includes large uncertainty. 
We show numerically that 200\% uncertainty for $a_{1}$ can be interpreted 
as about 2\% extra uncertainty for $a_{0}$, which should be added to the 
original uncertainty of $a_{0}$, and this result confirms our
expectations from the 
correlation.  This extra uncertainty makes the CP sensitivity worse, 
especially within the large $|U_{e3}|$ region.  
A shorter baseline and lower energy option can
avoid this disadvantage if the detection threshold can be lowered.  On
the contrary, if small $|U_{e3}|$ is established, then a long-baseline 
and high-energy option may be better than a shorter baseline
and lower energy, because of the statistics.

We made some comments in answer to questions about the statistics which
we used.  We consider the fact that the best-fit parameter suggested by
the experiments is not always distributed only around the parameter
chosen by nature. Therefore, to discuss the feasibility of observing the CP
violation effect, we need to consider this fact. 
This led us to use the concept, the power of test.

In this study, we do not consider systematic uncertainty. Of course, in
order to
optimize the experimental configurations, it is necessary to take
account of systematic error.  However, in any case, we can conclude
that we should regard the uncertainty of the Earth's matter as a more
severe problem than that has so far been assumed. We need to be much more
conservative when estimating the matter effect.

\begin{acknowledgments}
The authors thank the Yukawa Institute for Theoretical Physics at Kyoto
University.
Discussions during the YITP workshop YITP-W-02-05 on
``Flavor mixing, CP violation and Origin of matter''
were useful for completing this work.
 T. O. would like to thank W. Winter for his useful comments.
The work of J. S. is supported in part by Grants-in-Aid for 
Scientific Research from the Ministry of Education, Science, Sports, 
and Culture of Japan, No. 14740168, No. 14039209 and No. 14046217.
The English used in this manuscript was revised by K. Miller (Royal
 English Language Centre, Fukuoka, Japan).
\end{acknowledgments}


\end{document}